\documentclass[aps,twocolumn,showpacs,pra,twoside,amssymb,amsmath]{revtex4}
\usepackage{amssymb}
\usepackage{graphicx,color}
\usepackage{amsmath}
\usepackage{colordvi}
\usepackage{bbm}
\usepackage{amsmath}

\newcommand{\ket}[1]{|#1\rangle}

\newcommand{\tr}{{\rm Tr}}
\newcommand{\mX}{{\mathcal X}}
\newcommand{\mJ}{{\mathcal J}}

\begin{document}

\title{Quantum discord of two-qubit  $X$-states}
\author{Qing Chen$^{1,2}$}
\email{cqtcq@nus.edu.sg}
\author{Chengjie Zhang$^{1}$}
\author{Sixia Yu$^{1,2}$}
\author{X.X. Yi$^{1,3}$}
\author{C.H. Oh$^{1}$}
\email{phyohch@nus.edu.sg}
\affiliation{$^1$Department of P hysics and Centre for Quantum Technologies, National University of Singapore, 117543, Singapore\\
$^2$Hefei National Laboratory for Physical Sciences at Microscale and Department of Modern Physics, \\
University of Science and Technology of China, Hefei, Anhui 230026, China \\
$^3$ School of Physics and Optoelectronic Technology, Dalian University of Technology, Dalian 116024, China
}

\begin{abstract}
Quantum discord provides a measure for quantifying quantum correlations beyond entanglement and is very hard to compute even for two-qubit states because of the minimization over all possible measurements. Recently a simple algorithm to evaluate the quantum discord for two-qubit $X$-states is proposed by Ali, Rau and Alber [Phys. Rev. A \textbf{81}, 042105 (2010)] with minimization taken over only a few cases. Here we shall at first identify a class of $X$-states, whose quantum discord can be evaluated analytically without any minimization, for which their algorithm is valid, and also identify a family of $X$-states for which their algorithm fails. And then we demonstrate that this special family of $X$-states provides furthermore an explicit example for the inequivalence between the minimization over positive operator-valued measures and that over von Neumann measurements.
\end{abstract}


\maketitle

It is believed that entanglement is an essential resource in quantum computation and communication.
However, there are also quantum tasks that display the quantum advantage without entanglement,
for example, the deterministic quantum computation with one qubit \cite{DQC1}.
Therefore, there is a need to consider quantum correlations beyond entanglement \cite{Datta}. The quantum discord  has been introduced by Ollivier and Zurek \cite{discord1} and
independently by Henderson and Vedral \cite{discord2} to quantify quantum correlations. Recently the quantum discord has attracted much interest in quantum information theory \cite{Modi,Devi,MID,merge,phase} such as its relation  with the complete positivity \cite{CP} and local broadcasting of the state \cite{broadcast}. Furthermore, both Markovian and non-Markovian dynamics of quantum discord have been analyzed not only in theory but also in experiments \cite{almost,Markovian1,Markovian2,Markovian3,Markovian4}.
Whether the quantum discord can be more robust against decoherence \cite{almost,Markovian1} than entanglement or not \cite{zhangjs} is still an open problem \cite{zhangjs}.

The quantum discord is always nonnegative \cite{discord1}. States with vanishing quantum discord are relatively well understood and necessary and sufficient conditions are obtained to detect nonzero quantum discord \cite{condition1} as well as nonlinear witnesses have been proposed both for a given state \cite{Rahimi} and an unknown state \cite{zhang}.  Unfortunately almost all quantum states have nonzero quantum discords \cite{almost}, which  are notoriously difficult to compute because of the minimization over all possible positive operator-valued measures (POVMs) or von Neumann measurements. In addition to a few analytical results including the Bell-diagonal states  \cite{bell1}, rank-2 states \cite{rank2}, and gaussian states \cite{gauss},  a thorough numerical calculation \cite{max} has also been carried out in the case of von Neumann measurements for two-qubit states.

For an important family of two-qubit states, the so called $X$-states \cite{yut}, an algorithm has been proposed to calculate their quantum discord with minimization taken over only a few simple cases \cite{X2}, which is unfortunately impeded by a counter example \cite{Lu}. In this paper we shall at first identify a vast class of $X$-states, whose quantum discord can be evaluated analytically without any minimization at all, for which their algorithm is valid, and also identify a family of $X$-states $\mX_m$, the so-called maximally discordant mixed states \cite{max}, for which the above mentioned algorithm fails. And then for this family of $X$-states $\mX_m$ we construct a POVM showing that the quantum discord obtained by minimization over all POVMs is strictly smaller than that over all possible von Neumann measurements.

For  a given quantum state $\varrho$ of a composite system $AB$ the total amount of correlations, including classical and quantum correlations, is quantified by the quantum mutual information
$\mathcal{I}(\rho)=S(\varrho_A)+S(\varrho_B)-S(\varrho)$ where $S(\varrho)=-\mathrm{Tr}(\varrho\log_2\varrho)$ denotes the von Neumann entropy and $\varrho_A$, $\varrho_B$ are reduced density matrices for subsystem $A$, $B$ respectively.
An alternative version of the mutual information can be defined as
\begin{equation}\label{ci}
    \mathcal{\tilde J}_A(\varrho)=S(\varrho_B)-\min_{\{E_k^A\}}\sum_k p_k S(\varrho_{B|k})
\end{equation}
where the minimum is taken over all possible POVMs
$\{E_k^A\}$ on subsystem $A$
with $p_k=\mathrm{Tr}(E_k^A\varrho)$ and $\varrho_{B|k}=\tr_A(E_k^A\varrho)/p_k$.
Since $\mathcal{\tilde J}_A(\varrho)$ quantifies the classical correlation, the difference
\cite{discord2}
 \begin{eqnarray}\label{}
 \tilde {D}_A(\varrho)&=& \mathcal{I}(\rho)- \mathcal{\tilde J}_A(\varrho) 
\end{eqnarray}
defines the quantum discord that quantifies the quantum correlation.
Also the minimum in Eq.(\ref{ci}) can be  taken over all von Neumann measurements \cite{discord1} and we denote the corresponding
classical correlation as $\mathcal{J}_A(\varrho)$ and quantum discord as $ D_A(\varrho)$, respectively.
Obviously $\tilde D_A(\varrho)\le D_A(\varrho)$ and it becomes an equality for some states such as Bell-diagonal states and a family of filtered $X$-states \cite{yu}.

The two-qubit $X$-state usually arises as the two-particle reduced density matrix in many physical systems possessing $z$-axis symmetry. In the computational basis $\{\ket{00},\ket{01},\ket{10},\ket{11}\}$ its density matrix
\begin{equation}\label{Eq:Xstate}
\mX=\left(\begin{array}{cccc} \varrho_{00}&0&0&\varrho_{03}\cr 0&\varrho_{11}&\varrho_{12}&0\cr
0&\varrho_{12}^*&\varrho_{22}&0\cr \varrho_{03}^*&0&0&\varrho_{33}\end{array}\right)
\end{equation}
has seven real parameters. Via local unitary transformations, which preserve the quantum discord, elements $\varrho_{03}$ and $\varrho_{12}$ can be brought into real numbers. Thus there are in fact only five real parameters, which can be conveniently taken as
\begin{eqnarray}
x&=&\varrho_{00}+\varrho_{11}-\varrho_{22}-\varrho_{33}=\tr(\sigma_z^A\mX),\nonumber \\
y&=&\varrho_{00}-\varrho_{11}+\varrho_{22}-\varrho_{33}=\tr(\sigma_z^B\mX),\nonumber \\
t&=&\varrho_{00}-\varrho_{11}-\varrho_{22}+\varrho_{33}=\tr(\sigma_z^A\sigma_z^B\mX),\label{pa}\\
s&=&2(\varrho_{12}+\varrho_{03})=\tr(\sigma_x^A\sigma_x^B\mX),\nonumber \\
u&=&2(\varrho_{12}-\varrho_{03})=\tr(\sigma_y^A\sigma_y^B\mX),\nonumber
\end{eqnarray}
where $\sigma_{x,y,z}^{A,B}$ are three standard Pauli matrices. The positivity requires $(1\pm t)^2\ge {(x\pm y)^2+(s\mp u)^2}$ with all five parameters taking values in the interval $[-1,1]$. Without loss of generality we shall assume $|s|\ge |u|$ in what follows because we can always change the sign of $\varrho_{03}$ by a local unitary transformation.

The quantum discords of some $X$-states have been numerically calculated \cite{Markovian2,Markovian3,X1}.
Most recently an algorithm \cite{X2} has been proposed to calculate quantum discord for all two-qubit $X$-states in which the minimization is taken over only a few simple cases  instead of  actually minimizing over all possible measurements.
 In the case of real $X$-state with $|s|\ge |u|$ the algorithm reads:
for $D_A(\mX)$
\begin{equation}\label{alg}
\mbox{ the optimal observable is either }\sigma_x^A\; \mbox{or}\; \sigma_z^A.
\end{equation}
Recently a counter example is found in \cite{Lu} for this algorithm. The following theorem identifies a region of parameters of $X$-state, whose  quantum discord can be evaluated analytically without any minimization, for which the above mentioned algorithm is valid.

{\bf Theorem } The optimal measurement for the quantum discord $D_A(\mX)$ and $\tilde {D}_A(\mX)$ of a real $X$-state $\mX$ with $|s|\ge|u|$  is
 i) $\sigma_z^A$ if
\begin{equation}\label{c1}
(|\varrho_{12}|+|\varrho_{03}|)^2\le (\varrho_{00}-\varrho_{11})(\varrho_{33}-\varrho_{22})
\end{equation}
and ii) $\sigma_x^A$ if $\left|\sqrt{\varrho_{00}\varrho_{33}}-\sqrt{\varrho_{11}\varrho_{22}}\right|\le |\varrho_{12}|+|\varrho_{03}|.$

{\bf Proof }
 Considering a general POVM $\{\mu_k(1+\vec n_k\vec \sigma^A)\}_{k=1}^K$ with $K \leq 4$ \cite{Ariano} made on the first qubit,
 where $\sum_k\mu_k=1$, $\vec n^2_k=1$, and $\sum_k\mu_k\vec n_k=0$,
  we obtain the outcome $k$ with probability $p_k=\mu_k(1+x n_{kz})$.
The second qubit is in the conditioned state
\begin{eqnarray}
\mX_{B|k}
=\frac {1+x n_{kz} + s n_{kx}\sigma_x^B + u n_{ky}\sigma_y^B+(y + n_{kz} t)\sigma_z^B}{2(1+x n_{kz})}.
\end{eqnarray}
By denoting $h(w)=-\frac{1+w}{2}\log_2\frac{1+w}{2}-\frac{1-w}{2}\log_2\frac{1-w}{2}$
and $\Delta_k=(1-n_{kz}^2)s^2+(y + n_{kz}t)^2$, we obtain
\begin{eqnarray}
\sum_{k}p_k S(\mX_{B|k})
\ge\sum_{k}{p_k} h\left(\frac{\mu_k \sqrt{\Delta_k}}{p_k}\right):=S_\mX,\label{ce}
\end{eqnarray}
where the inequality is due to the function $h(w)$ being a decreasing function for $w\ge 0$ and the equality can be attained by taking $n_{ky}=0, \forall k$.
Note that in the case of von Neumann measurements,
 since any observable $\sigma_n^A$ not in the $x$-$z$ plane the observable $\sigma^A_{n^\prime}$ with $\vec n^\prime=(\sqrt{1-n_z^2},0,n_z)$ will yield a smaller value, the optimal von Neumann measurement must lie in the $x$-$z$ plane.

Note that in the case of von Neumann measurements,
any observable $\sigma_n^A$ not in the $x$-$z$ plane the observable $\sigma^A_{n^\prime}$ with $\vec n^\prime=(\sqrt{1-n_z^2},0,n_z)$ will yield a smaller value.
Thus the optimal von Neumann measurement must lie in the $x$-$z$ plane.

By denoting $\lambda_{k\pm}=\mu_k(1+x n_{kz} \pm \sqrt{\Delta_k})/2$ and its marginal $\lambda_{k}=\lambda_{k+}+\lambda_{k-}=\mu_k(1+x n_{kz})$,
we obtain
\begin{equation}
S_\mX=\sum_k \lambda_{k} \log_2 {\lambda_{k}} - \sum_{k\pm} \lambda_{k\pm} \log_2 {\lambda_{k\pm}}.
\end{equation}
In the following we will find out the minimum of $S_\mX$ for
fixed $\{ \mu_1, ..., \mu_K \}$. Since $\sum_k\mu_k=1$ and $\mu_k\ge 0$ we can suppose without loss of generality that $\mu_K>0$. Because of condition $\sum_k\mu_k n_{kz}=0$ we can regard $n_{Kz}$ as a function of $K-1$ independent variables $n_{kz}:=n_k$ with $k=1,2,\ldots,K-1$. Therefore $S_\mX$ is a multivariable function of  $n_{k}$ with $k=1,...,K-1$ and its Hessian matrix, whose elements are
$W_{ij}=\frac{\partial^2 S_\mX}{\partial n_i \partial n_j}$, reads
\begin{widetext}
\begin{eqnarray}
\frac{W_{ij}}{\ln2} &=&\sum_{k=1}^K \frac{\lambda_k^i \lambda_k^j }{\lambda_{k}}
- \sum_{k=1,\pm}^K \frac{\lambda_{k\pm}^i \lambda_{k\pm}^j }{\lambda_{k\pm}}
- \sum_{k\pm} \lambda_{k\pm}^{ij}\ln \lambda_{k\pm} \nonumber \\
&=& - \sum_{k=1}^K   \frac{\Lambda_k^i\Lambda_k^j}
{\lambda_{k}\lambda_{k+}\lambda_{k-}}
- \frac{s^2(t^2-y^2-s^2)}{2}\left(\frac{\delta_{ij}\mu_i}{\sqrt{\Delta_i^3}}\ln\frac{\lambda_{i+}}{\lambda_{i-}}
+\frac{\mu_i\mu_j}{\mu_K\sqrt{\Delta_K^3}}\ln\frac{\lambda_{K+}}{\lambda_{K-}}  \right),
\end{eqnarray}

where $\lambda_{k\pm}^i = \frac{\partial \lambda_{k\pm}}{\partial n_i}$,
$\lambda_{k\pm}^{ij} = \frac{\partial^2 \lambda_{k\pm}}{\partial n_i \partial n_j}$, and $\Lambda_k^i=\lambda_{k-}\lambda_{k+}^{i}-\lambda_{k+}\lambda_{k-}^{i}$.
\end{widetext}
If $t^2\ge y^2+s^2$, which is equivalent to Eq.(\ref{c1}),
the matrix $W$ is always negative semidefinite, thus
the conditional entropy $S_\mX$ is a concave function of $n_{iz}$.
Therefore the minimum of $S_\mX$ is attained on the boundary, i.e., $n_{iz}=-1$ or $n_{iz}=1, \forall i$.
Thus for every given $\{ \mu_1, ..., \mu_K \}$, the optimal measurement is $\sigma_z^A$, which proves the first case of the Theorem.
For the proof of the second case, we refer to \cite{yu}. \hfill Q.E.D.

\begin{figure}
\begin{center}
\includegraphics[scale=0.84]{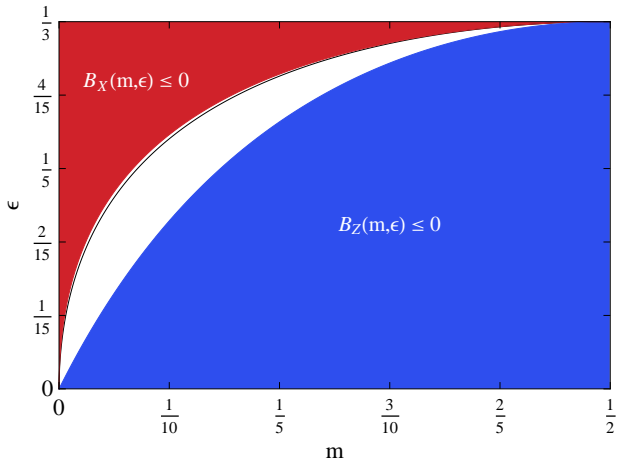}
\caption{ (Color online) The optimal observable for the state $\mX_3$ parameterized by $m$ and $\epsilon$ is i) $\sigma_x$ if $B_X(m,\epsilon)\leq 0$ (top red region) and ii) $\sigma_z$ if $B_Z(m,\epsilon)\leq 0$ (bottom blue region). The black curve corresponds the state $\mX_m$.
}
\label{ep-m}
\end{center}
\end{figure}

As the first example we consider the Bell-diagonal states for which we have $x=y=0$. If $|t|\ge |s|$ then we have case i) so that the optimal observable is $\sigma_z$ and if $|t|\le |s|$ then we have case ii) so that the optimal observable is $\sigma_x$ (recalling that we have assumed $|s|\ge |u|$), both reproduce the result in \cite{bell1}.
As the second example we consider a 2-parameter family of $X$-states
\begin{eqnarray}
\mX_3 = \left(
\begin{array}{cccc}
\epsilon/2 & 0 & 0 & \epsilon/2 \\
0 & (1-\epsilon)m & 0 & 0 \\
0 & 0 & (1-\epsilon)(1-m)& 0 \\
\epsilon/2 & 0 & 0 & \epsilon/2
\end{array}
\right),
\label{Eq:X0}
\end{eqnarray}
with $0\le \epsilon,2m\le 1$.
In this case we have $x=-y=(1-\epsilon)(2m-1)$, $s=-u=\epsilon$, and $t=2\epsilon-1$.
According to two cases in the Theorem the optimal observable is $\sigma_x$ or $\sigma_z$ in the case of $B_X(m,\epsilon)\leq 0$ or $B_Z(m,\epsilon)\leq 0$, respectively, where
\begin{eqnarray}
B_X(m,\epsilon) &=&\sqrt{m(1-m)} - \frac{\epsilon}{1-\epsilon},
 \label{Eq:XX} \\
B_Z(m,\epsilon)&=&\frac{\epsilon}{1-\epsilon}-2m(1-m).
\label{Eq:ZZ}
\end{eqnarray}
If $\epsilon\ge 1/3$ we have $B_X(m,\epsilon)\le 0$ and if $\epsilon\le 1/3$ we have illustrated in Fig.\ref{ep-m} those two regions (top red and bottom blue) of parameters for which the optimal observable is either $\sigma_x$ or $\sigma_z$  and in these regions the algorithm (\ref{alg}) is valid.

\begin{figure}
\begin{center}
\includegraphics[scale=0.44]{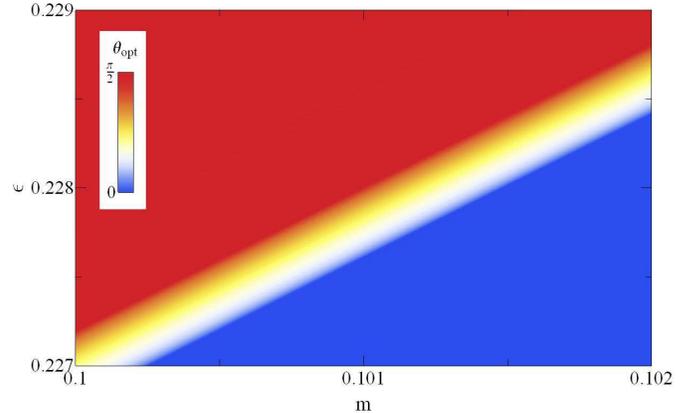}
\caption{(Color online) The optimal observable $\sigma_x\sin\theta_{opt}+\sigma_z\cos\theta_{opt}$ for the quantum discord $D_A(\mX_3)$.
}
\label{comput1}
\end{center}
\end{figure}

However there also exists a region (white) of parameters about which our theorem does not say anything.
To find out the quantum discord $D_A(\mX_3)$ in this region we have to do numerical calculations.
Taking into consideration the eigenvalues
$\lambda(\mX_3)=\{ \epsilon, (1-\epsilon)m, (1-\epsilon)(1-m),0  \}$ and
$\lambda(\tr_B\mX_3)=\{(1\pm x)/2\}$ we
obtain the value
\begin{eqnarray}
D_\theta = S_{\mX_3}(\cos\theta)+h(x)-h(t)-(1-\epsilon)h(2m-1)
\end{eqnarray}
after the  measurement of the observable along an arbitrary direction $\vec n=(\sin\theta,0,\cos\theta)$ in the $x$-$z$ plane. Here $S_{\mX_3}(\cos\theta)$ is defined in Eq.(\ref{ce}) with $n_z=\cos\theta$. The minimization of $D_\theta$ over all possible angles gives the quantum discord, i.e., $D_A(\mX_3)=\min_\theta D_\theta:=D_{\theta_{opt}}$. We note that $D_0=\epsilon$ and $S_{\mX_3}(0)=h(\sqrt{y^2+\epsilon^2})$.

A detailed numerical search for $D_A(\mX_3)$ in the ranges $m\in [0.1,0.102]$ and $\epsilon\in[0.227,0.229]$ has been carried out with results shown in Fig.2. The parameters for which the optimal observable is $\sigma_x$ or $\sigma_z$ are highlighted in red (top) or blue (bottom) respectively. However these two regions are separated by an intermediate region for which the optimal observable is neither $\sigma_x$ nor $\sigma_z$. As a result, though valid for most of the parameters, the algorithm (\ref{alg}) fails for a region with finite measure. Moreover our numerical search shows that there are about $0.05 \%$ of $5 \times 10^{7}$  randomly chosen $X$-states satisfying $s=u$ that violate the algorithm (\ref{alg}). In our search the difference between the true value of $D_A(\mX)$ and $\min\{D_0,D_{\frac\pi2}\}$ can reach as high as $0.0029$ in the case of $x=-0.8812, y=0.9407, s=0.2898,t=-0.9383$.

Especially we have considered in more details a subfamily of $X$-state $\mX_m$ of $\mX_3$ with $\epsilon$ determined by the condition $D_0=D_{\frac\pi2}$ or more explicitly
\begin{equation}
h(\sqrt{x^2+\epsilon^2})+h(x)-h(t)-(1-\epsilon) h(2m-1)=\epsilon
\end{equation}
recalling that $x=(1-\epsilon)(2m-1)$ and $t=2\epsilon-1$. This special family of $X$-states $\mX_m$ is exactly a family of so-called maximal discordant states investigated in \cite{max}. As a function of $m$ the solution $\epsilon(m)$ of the above equation is plotted as a black curve in Fig.1.

\begin{figure}
\begin{center}
\includegraphics[scale=0.8]{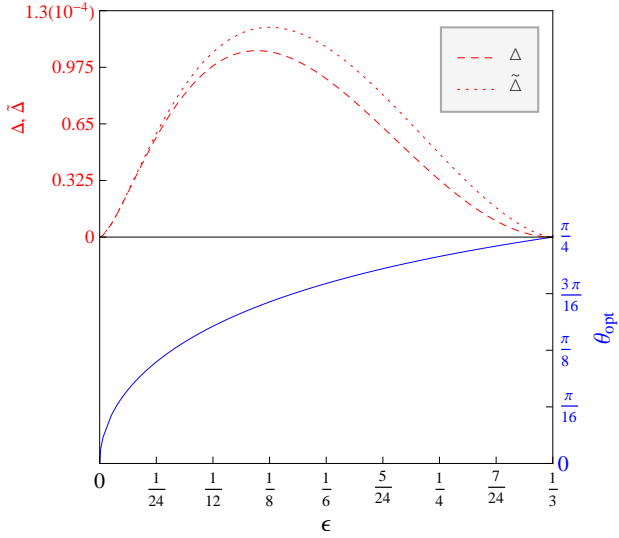}
\caption{(Color online) As the functions of $\epsilon$, the optimal angle $\theta_{opt}$ for $D_A(\mX_m)$, the difference $\Delta=\epsilon-D_A(\mX_m)$, and
the difference $\tilde \Delta=\epsilon-\tilde D^{upper}_A(\mX_m)$ are plotted in solid blue, dashed red and dotted red curves, respectively.}
\label{delta-theta-ep}
\end{center}
\end{figure}

\begin{figure}
\begin{center}
\includegraphics[scale=0.8]{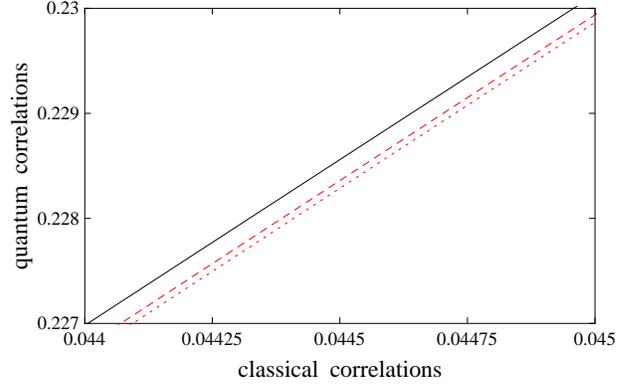}
\caption{(Color online) The $\mJ$-$D$ diagram for $\mX_m$ with its quantum discord taken as $D_0$ (as in \cite{max}) (solid black line), $D_{\theta_{opt}}$ (dashed red line), and $\tilde D^{upper}_A(\mX_m)$ (dotted red line).
}
\label{q-c}
\end{center}
\end{figure}

In Fig.\ref{delta-theta-ep} the optimal angle $\theta_{opt}$ (solid blue line) and
the difference $\Delta=D_0-D_{\theta_{opt}}$ (dashed red line) are plotted as functions of $\epsilon$.
As $\epsilon (m)$ increases, $\Delta$ increases at first from 0 to its maximum about $1.07 \times 10^{-4}$ at $\epsilon \simeq 0.115699$ and then decreases to zero.
Except at the end points
$(m,\epsilon)=(0,0), (1/2,1/3)$ for which any observable is optimal, we have $\Delta>0$ which means that $\sigma_x$ and $\sigma_z$ are not optimal.
Most interestingly the optimal angle $\theta_{opt}$ takes values in $[0,\pi/4]$ continuously.
This fact strengthens the theorem proposed in Ref. \cite{Lu}, i.e., it is impossible to find a
universal finite set of optimal von Neumann measurements even for the real $X$-states.

In Ref. \cite{max}, by maximizing the quantum discord for given classical correlations over all states $\mX_3$, whose quantum discord is taken to be $\min\{D_0, D_{\frac\pi2}\}$, the state $\mX_m$ turns out to be a family of maximally discordant mixed state that lies on the boundary of the $\mJ$-$D$ diagram of the classical-correlation vs quantum-discord.
Since $D_A(\mX_m)< \min\{D_0, D_{\frac\pi2}\}$ except at the endpoints, the $\mJ$-$D$ diagram for $\mX_m$ is shifted right-downward as shown in Fig. \ref{q-c}. Thus whether the state $\mX_m$ is still on the boundary or not needs further numerical calculation to substantiate.

Finally let us consider the quantum discord $\tilde D_A(\mX_m)$ obtained by minimization over all possible POVMs. For each given state $\mX_m$ there is an optimal angle $\theta_{opt}$ for $D_A(\mX_m)$ obtained by von Neumann measurements and we perform a 3-outcome POVM
$\{\mu_k(1+\vec n_k\vec \sigma^A)\}_{k=1}^3$ made on the first qubit on the subsystem $A$,
where
\begin{eqnarray}
\mu_1=\frac{\cos\theta_{opt}}{1+\cos\theta_{opt}}, && \vec n_1 = \{0,0,-1\}, \nonumber  \\
\mu_2=\frac{1}{2(1+\cos\theta_{opt})}, &&  \vec n_2 = \{\sin\theta_{opt},0, \cos\theta_{opt} \},  \\
\mu_3=\frac{1}{2(1+\cos\theta_{opt})}, &&  \vec n_3 = \{-\sin\theta_{opt},0,\cos\theta_{opt} \}. \nonumber
\end{eqnarray}
We denote by $\tilde D^{upper}_A(\mX_m)$ the corresponding suboptimal value for the quantum discord and obviously $\tilde D_A(\mX_m)\le \tilde D^{upper}_A(\mX_m)$.
The difference  $\tilde \Delta= D_0 - \tilde D^{upper}_A(\mX_m)$ is shown in Fig. \ref{delta-theta-ep} as a dotted red line and we have $\Delta<\tilde\Delta$ (except at end points), which means that the $\mJ$-$D$ diagram for $\mX_m$ must be shifted further right-downward as shown in Fig. \ref{q-c} by the dotted red line. The boundaries
of the $\mJ$-$D$ diagrams for POVMs and von Neumann measurements would be different if $\mX_m$ were still the maximally discordant states.

To summarize, we have presented some positive results as well as negative results on the quantum discord of $X$-state.
We have identified a vast class of $X$-states whose quantum discords can be evaluated analytically and also a family of maximally discordant mixed states $\mX_m$
that invalidate the algorithm \cite{X2}. If the state $\mX_m$ were still on the boundary of the $\mJ$-$D$ diagram \cite{max}  then this part of boundary would be shifted right-downward and even further for POVMs.
Thus the state $\mX_m$ provides
an explicit example for the inequivalence between the minimization over POVMs and that over von Neumann measurements for X-states.
Recently, more examples \cite{Galve} have been given.

This work is supported by National Research Foundation and Ministry of Education, Singapore (Grant No. WBS: R-710-000-008-271)
and NNSF of China (Grants No. 11075227 and No. 10935010).

\end{document}